# Polarimetric imaging for the detection of synthetic models of SARS-CoV-2: a proof of concept

**Emilio Gomez-Gonzalez[1,2*], Olga Muñoz[3], Juan Carlos Gomez-Martin[3], Jesus Aceituno-Castro[3,4], Beatriz Fernandez-Muñoz[6], Jose Manuel Navas-Garcia[7], Alejandro Barriga-Rivera[1,8], Isabel Fernandez-Lizaranzu[1,2], Francisco Javier Munoz-Gonzalez[1], Ruben Parrilla-Giraldez[5], Desiree Requena-Lancharro[1], Pedro Gil-Gamboa[1], José Luis Ramos[3], Cristina Rosell-Valle[6], Carmen Gomez-Gonzalez[10], Maria Martin-Lopez[6], Maria Isabel Relimpio-Lopez[11,12,13], Manuel A. Perales-Esteve[1,9], Antonio Puppo-Moreno[2,10], Francisco Jose Garcia-Cozar[14,15], Lucia Olvera-Collantes[14,15], Silvia de los Santos-Trigo[16], Emilia Gomez[17], Rosario Sanchez-Pernaute[6], Javier Padillo-Ruiz[2], and Javier Marquez-Rivas[1,2,18,19]**

[1] Group of Interdisciplinary Physics, Department of Applied Physics III at the ETSI Engineering School, Universidad de Sevilla, 41092 Seville, Spain.
[2] Institute of Biomedicine of Seville, Spain.
[3] Cosmic Dust Laboratory, Instituto de Astrofísica de Andalucía, CSIC; 18008 Granada, Spain.
[4] Centro Astronomico Hispano Alemán, 04550 Almeria, Spain
[5] Technology and Innovation Centre, Universidad de Sevilla; 41012 Sevilla, Spain.
[6] Unidad de Producción y Reprogramación Celular, Red Andaluza de Diseño y Traslación de Terapias Avanzadas, Consejería de Salud y Familias, Junta de Andalucía; 41092 Sevilla, Spain.
[7] EOD-CBRN Group, Spanish National Police, 41011 Sevilla, Spain.
[8] School of Biomedical Engineering, The University of Sydney, NSW 2006, Australia.
[9] Department of Electronic Engineering at the ETSI Engineering School, Universidad de Sevilla, 41092 Seville, Spain.
[10] Service of Intensive Care, University Hospital 'Virgen del Rocio'; 41013 Sevilla, Spain.
[11] Department of General Surgery, College of Medicine, Universidad de Sevilla, 41009 Seville, Spain,
[12] Department of Ophthalmology, University Hospital 'Virgen Macarena', 41009 Sevilla, Spain
[13] OftaRed, Institute of Health 'Carlos III'; 28029 Madrid, Spain.
[14] Department of Biomedicine, Biotechnology and Public Health, University of Cadiz, 11003 Cadiz, Spain
[15] Instituto de Investigación e Innovación Biomedica de Cádiz (INIBICA); 11009 Cadiz, Spain.
[16] Corporación Tecnológica de Andalucía, 41092 Sevilla, Spain.
[17] Joint Research Centre, European Commission; 41092 Sevilla, Spain.
[18] Service of Neurosurgery, University Hospital 'Virgen del Rocío', 41013 Sevilla, Spain
[19] Centre for Advanced Neurology, 41013 Sevilla, Spain.

*Corresponding author.
E-mail: egomez@us.es

## Abstract

Objective: To conduct a proof-of-concept study of the detection of two synthetic models of severe acute respiratory syndrome coronavirus 2 (SARS-CoV-2) using polarimetric imaging. Methods: Two SARS-CoV-2 models were prepared as engineered lentiviruses pseudotyped with the G protein of the vesicular stomatitis virus, and with the characteristic Spike protein of SARS-CoV-2. Samples were preparations in two biofluids (saline solution and artificial saliva), in four concentrations, and deposited as 5-μL droplets on a supporting plate. The angles of maximal degree of linear polarization (DLP) of light diffusely scattered from dry residues were determined using Mueller polarimetry of 87 samples at 405 nm and 514 nm. A polarimetric camera was used for simultaneous imaging of several samples under 380-420 nm illumination at angles similar to those of maximal DLP. A per-pixel image analysis included quantification and combination of polarization feature descriptors in other 475 samples. Results: The angles (from sample surface) of maximal DLP were 3 degrees for 405 nm and 6 degrees for 514 nm. Similar viral particles that differ only in the characteristic spike protein of the SARS-CoV-2, their corresponding negative controls, fluids, and the sample holder were discerned from polarimetric image analysis at 10-degree and 15-degree configurations. Conclusion: Polarimetric imaging in the visible spectrum has the potential for non-contact, reagent-free detection of viruses in multiple dry fluid residues simultaneously. Further analysis including real SARS-CoV-2 in human samples -particularly, fresh saliva- are required. Significance: Polarimetric imaging under visible light could contribute to fast, cost-effective screening of SARS-CoV-2 and other pathogens.

## 1. Introduction

The wide spread of coronavirus disease 2019, known as COVID-19, has mobilized unprecedented economic resources for the quest of finding new ways of fighting the propagation of infectious diseases. A vast amount of strategies have been implemented from almost all fields of research, including molecular biology [1, 2], pharmacology [3-5], genetics [6, 7], and medical imaging [8-11] and artificial intelligence [12], among others, with varying levels of success [13]. Three main lines of action can be highlighted: prevention [14], treatment[15], and detection (both for screening and diagnostic purposes) [16]. Severe social distancing measures have been imposed worldwide to reduce the propagation of the so-called severe acute respiratory coronavirus 2 (SARS-CoV-2). With the arrival of a collection of vaccines [17, 18], some of these measures have been eased in those territories with high vaccination rates [19-22]. However, an important surge in the number of new infections has been reported following a period after inoculation of the first vaccines [23, 24]. In addition, population that have not had full access to preventive medication is still large. Therefore, the need for testing and monitoring for the spread of the SARS-CoV-2 remains.

A family of molecular testing devices have recently emerged to identify past [25, 26] and on-going infections [25, 27]. While debate still continues about the establishment of a gold standard technique for the diagnosis of COVID-19, polymerase chain reaction (PCR) remains as the accepted benchmark. However, rapid tests have enabled screening at the point of care within minutes [28], although with reduced accuracy.

Optical techniques are also explored for the detection of a variety of pathogens. They range from the detection of the Zika virus in mosquitoes fed with infected human blood using infra-red spectroscopy [29] to the detection of hepatitis B and C viruses [30] and Dengue [31] in human samples using Raman and fluorescence spectroscopy. Most recent efforts aim to detect SARS-CoV-2 models [32] and the SARS-CoV-2 virus [33-36].

These techniques are opening new possibilities for rapid concurrent mass screening of contagious diseases and have the potential to re-shape the screening approaches currently in place at transport hubs such international airports. However, the optical information contained in biological samples has not been fully exploited. For example, Mueller polarimetry [37, 38], a technique employed to characterize the sample photo-polarimetric characteristics can be used for the identification of the optical signature of biological structures [39-41]. In fact, Mueller polarimetry is used in astrobiology in the search for biological molecules (biosignatures) outside the Earth [42-44]. Recent works highlight the strong potential of the analysis of the vectorial properties of light beams interacting with samples for many biomedical applications [45], including the detection of viruses [46, 47]. These approaches range from physics-based methods to artificial intelligence (e.g., machine learning) algorithms. Notably, computational simulations show the potential of the evaluation of changes of circular polarization under ultraviolet light to detect and discern viral model particles with the characteristic Spike protein of the SARS-CoV-2 [48].

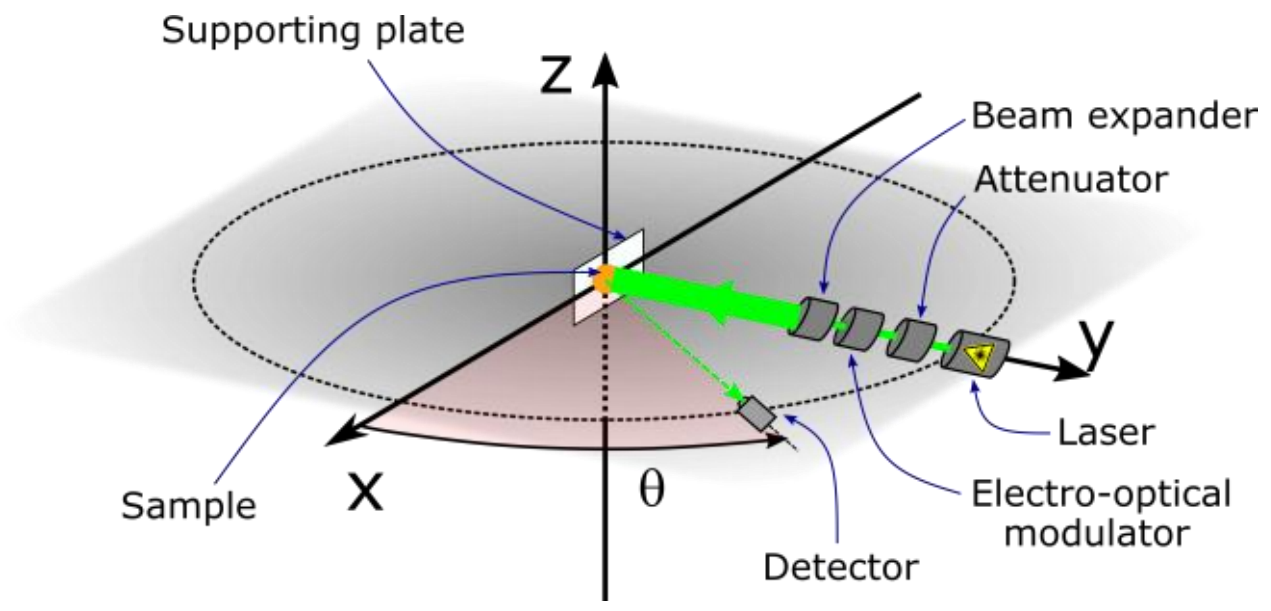

Fig. 1. Schematic of the optical setup used to determine the viewing angle (θ) that maximizes the degree of linear polarization of the diffusely reflected light. A continuous-wave laser illuminated the sample. The light beam passed through an optical attenuator to reduce its intensity, and an electro-optical modulator modified its polarization state. A beam expander provided homogenous illumination of the sample.

Here we have examined the use of polarimetric imaging under visible light for the rapid identification of engineered viral particles (two laboratory models of the SARS-CoV-2, with and without its Spike protein) deposited as fluid preparations on surfaces in an experimental proof-of-concept study. Firstly, we use the Cosmic Dust Laboratory (CODULAB) instrument [49] for obtaining the Mueller matrix of the light reflected by the surfaces of interest as a function of the viewing angle θ (Fig. 1). This first set of measurements spans a broad range of angles. Thus, they provide the optimum observing geometry for polarimetric imaging. Secondly, based on the previous results, we designed a polarimetric imaging set-up (Fig. 2) for simultaneous imaging of several samples placed on a supporting plate. Polarimetric imaging shows a strong potential for detection of elements in compound backgrounds [50] and classification of materials [51], although registered images are difficult to analyze because of the complex nature of physical processes involved and their multiplicative random speckle noise [52]. Recent approaches for their analysis include the use of deep learning techniques [53].

In this work, samples (i.e., droplets) can be easily segmented against the background, and a per-pixel image processing methodology has been applied. Several quantitative descriptors based on the measured polarimetric features are defined, calculated at the per-pixel level and later integrated at the per-sample level. This is the same procedure employed by authors for the analysis of hyperspectral images of the same type of samples which allowed for detecting and characterizing those ones with viral content in similar concentrations [32] [34].

## 2. Materials and methods

We analyze fluid samples containing two different types of synthetic models of the SARS-CoV-2, namely, lentiviral particles pseudotyped with the G protein of the vesicular stomatitis virus (G-LP), and with the characteristic Spike (S) protein of the SARS-CoV-2 (S-LP). Both models are ribonucleic acid viruses of similar size and shape as the SARS-CoV-2, also having a double lipid capsid, but different in the presence of the Spike protein on their surface. Samples were prepared in two different types of biofluids, a phosphate buffered solution (PBS), i.e., a type of saline solution, and artificial saliva (AS), much more similar to natural saliva. In each fluid, four levels of concentration (i.e., viral load) were tested. Negative controls included solutions of viral culture media (without viruses) in the same concentrations and pure fluids.

Optical set-ups in this study include i) The CODULAB instrument, to measure the reflection Mueller matrix of dry samples deposited on a supporting plate as function of the reflection angle ($\theta_R$) or its complementary the so called viewing angle ($\theta = 90° - \theta_R$) using as light sources two visible wavelength diode lasers (Fig. 1), and ii) a specific polarimetric imaging configuration under band-pass filtered halogen illumination (Fig. 2). Numerical descriptors of polarimetric features were then computed and combined in a per-pixel image processing procedure.

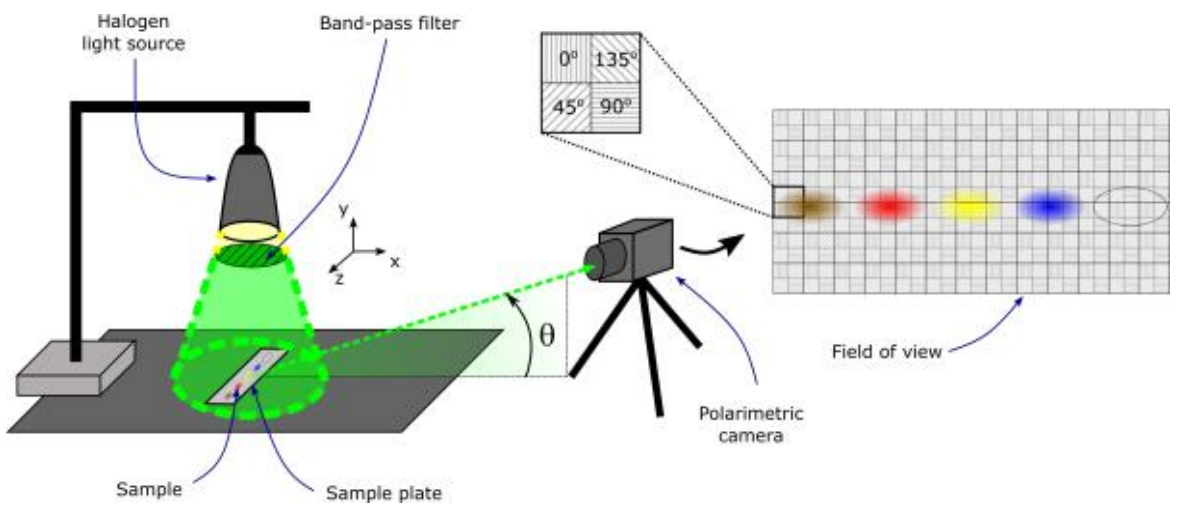

Fig. 2. Schematic of the setup used for polarimetric imaging. Samples were deposited as a row of four fluid droplets onto a supporting plate. They contained preparations of VSV-G pseudotyped lentiviral particles (brown area), SARS-CoV-2 Spike pseudotyped lentiviral particles (red area) and the corresponding culture medium (negative control, in the same concentration, yellow area) and the pure fluid (blue area). In addition, an equivalent region of the plate (dotted ellipse) was defined as background. The samples were illuminated with a halogen light source with a band-pass filter between 380 nm and 420 nm. The polarimetric camera was placed at a working distance of 90 mm from the sample, with the viewing axis in θ = 10o and θ = 15o configurations. Raw images were later demosaiced into independent I0, I90, I45 and I135 images (corresponding to each registered direction of linear polarization) for further processing.

### *2.1 Lentiviral particles pseudotyped with the G protein of the vesicular stomatitis virus (G-LP).*

Lentiviral particles pseudotyped with protein G of the vesicular stomatitis virus (VSV) were produced as described in a previous publication [32]. Briefly, human embryonic kidney (HEK) 293 cells were transfected with lentiviral plasmids encoding ZsGreen, the viral envelope VSV-G protein, and the Tat, Gag-Pol and Rev genes. These cells were then cultured in Gibco Dulbecco's Modified Eagle Medium (DMEM) and incubated during 48 hours at 37ºC and 5% $CO_2$. The culture medium was supplemented with 10% bovine serum (Biowest, Nuaillé, France), and the antibiotics penicillin ($100\ UI \cdot mL^{-1}$) and streptomycin ($100\ \mu g \cdot mL^{-1}$) (MilliporeSigma, Missouri, USA) were added to prevent bacterial infections. Lentiviral particles were precipitated using Lenti-X reagent (Takara Bio Inc, Shiga, Japan). Following centrifugation, the supernatant was removed and the concentrate was frozen at -80 ºC.

### *2.2 Lentiviral particles pseudotyped with the spike protein of the SARS-CoV-2 (S-LP).*

Similarly, lentiviral particles were pseudotyped with the Spike protein of the SARS-CoV-2 as described in another publication [34]. These engineered viral particles exhibit the molecular structure of the SARS-CoV-2 capsid characterized by the protuberant Spike protein that appears inserted on the surface of the virions. The preparation methodology was similar to that described above. Note that in this case, HEK 293 cells were transfected with a plasmid encoding the SARS-CoV-2 spike protein instead [54]. The preparation was cultured, precipitated and stocked as described previously for the VSV-G pseudotyped lentiviral particles.

### *2.3 Controls.*

Negative control samples were prepared following the same methodology described for pseudotyping lentiviral particles without transfecting the HEK cells, that is, the cells were cultured without adding plasmids and, therefore without viral particles. Incubation, precipitation and stocking was performed as described above. Resuspension of the media thus prepared was carried out to achieve the same titer as for the viral preparations in both media, PBS and AS. In addition, PBS and AS alone were also used as controls.

### *2.4 Sample preparation.*

On the day of the experiment, an aliquot was defrosted to prepare an initial stock with a titer of $20 \cdot 10^3$ transducing units $(TU) \cdot \mu L^{-1}$. The aliquot was re-suspended in two different media: phosphate buffered saline (MilliporeSigma, Missouri, USA) and 1700–0305 artificial saliva (Pickering Laboratories, California, USA). Serial dilutions were then prepared to achieve a final concentration of $4000\ TU \cdot \mu L^{-1}$, 3000 TU·µL$^{-1}$, 1500 TU·µL$^{-1}$ and 800 TU·µL$^{-1}$. As detailed in a previous paper [32], this lowest value corresponds approximately to $8\cdot10^{8}$ copies·mL$^{-1}$, i.e., to the viral load of potential 'super-spreaders' of COVID-19. Note that non-transfected samples were prepared in the same way to generate the controls. A total of 467 fluid droplets were prepared (87 for point polarimetry and 380 for imaging polarimetry).

## *2.5 Point Mueller polarimetry.*

CODULAB [49, 55] is designed to measure the full scattering Mueller matrix of clouds of cosmic dust analogues and atmospheric aerosols as function of the scattering angle. Here CODULAB is adapted to measure the reflection Mueller matrix, **R**, as function of the viewing angle, θ, of dry samples deposited on a supporting plate. The surfaces are illuminated with diode lasers emitting 10 mW at λ = 405 nm or λ = 514 nm orthogonally to the supporting plate (Fig. 1). Assuming homogeneous illumination on a 5 mm-diameter disk, the corresponding irradiance on the samples is about 509 W·m$^{-2}$. Electro-optic modulation of the incident beam combined with lock-in detection enables concurrent determination of the elements $R_{ij}$ (θ, λ) of the 4x4 reflection matrix, **R**(θ, λ). The modulated laser beam is reflected by the residue located on the supporting plate, which is mounted on a x-y rotating table to control the illumination geometry. To ensure homogenous illumination over the entire dry-residue, a 5-10x zoom Galilean beam expander with magnification set to 7x is positioned between the modulator and the sample. The reflected light is collected by a photomultiplier tube detector (9828A Electron tubes). This detector moves along a 1- diameter ring in steps of 1 or 5 degrees spanning the viewing angle range from 0 to -70°, with a blind region between 83° and -87°. The detector is located 62 cm away from the surface under study. For each data point at a given reflexion angle, 1000 measurements are conducted in about 1 s. Therefore, one single data point is the average of 1000 measurements. The error of one data point is the standard deviation of the series of 1000 measurements. A diaphragm is located at the detector to control solid angle subtended by the detector window. Therefore, each data point can be considered as an average value of the full dry-residue. These measurements are labelled as "point measurements" throughout the paper.

Initially, we investigated if any the $R_{ij}$ (θ, 405 nm) elements could help in discriminating clean surfaces from those covered with usual surface contaminants i.e., home dust and (artificial) saliva. A wide variety of surface materials (polymethyl methacrylate, polytetrafluoroethylene, steel, aluminum, cardboard) was tested. Different orientations of the sample holder with respect to the incident laser light were also tested. The optimum illumination geometry is found for the laser light incident perpendicular to the supporting plate. Subsequently, we tested if the $-R_{21}(\theta, \lambda)/R_{11}(\theta, \lambda)$ curve at 405nm and 514 nm could discriminate samples with viral particles from their negative controls. The $-R_{21}(\theta, \lambda)/R_{11}(\theta, \lambda)$ ratio is, for unpolarized incident light, equal to the degree of linear polarization, hereafter DLP [49]. In this set of measurements, the samples were placed on a polytetrafluoroethylene (PTFE) sheet (BSH, Seville, Spain), 1 mm in thickness, and approximately 22 mm × 22 mm of surface.

For each fluid, preparation, and concentration, a 5-µL droplet was deposited onto a holder plate. Upon complete evaporation of the water content only dry residues remained, and the sample plate was positioned vertically. A total of 87 fluid droplets were analyzed with this set-up (Table I). They included 16 samples of G-LP in PBS, 16 samples of S-LP in PBS and 8 samples of S-LP in artificial saliva at four different concentrations, their corresponding negative controls, and droplets of pure fluid. Measurements were carried out in a dark room with average temperature of 22ºC and humidity of 55%.

TABLE I
Number of samples used for Point Mueller polarimetry

| Sample type | Concentrations | | | | | | | |
|---|---|---|---|---|---|---|---|---|
| | C1 | | C2 | | C3 | | C4 | |
| | P | N | P | N | P | N | P | N |
| G-LP in PBS | 4 | 4 | 4 | 4 | 4 | 4 | 4 | 4 |
| S-LP in PBS | 4 | 4 | 4 | 4 | 4 | 4 | 4 | 4 |
| S-LP in AS | 2 | 1 | 2 | 1 | 2 | 1 | 2 | 1 |
| AS | 11 | | | | | | | |
| BG | 5 | | | | | | | |

Table I. Distribution of the 92 samples (87 fluid droplets and 5 droplet-equivalent background areas) used for point Mueller polarimetry. Concentrations were C1 = 800 TU·µL$^{-1}$, C2 = 1500 TU·µL$^{-1}$, C3 = 3000 TU·µL$^{-1}$ and C4 = 4000 TU·µL$^{-1}$. G-LP = Lentiviral particles pseudotyped with the G protein of the vesicular stomatitis virus. S-LP = Lentiviral particles pseudotyped with the spike protein of the SARS-CoV-2. P = positive sample (solution with viral particles and culture medium), N= negative sample (solution with culture medium). PBS = phosphate buffered solution, AS = artificial saliva, BG = background (supporting plate).

## *2.6 Polarimetric imaging.*

Based on the results obtained as described in section E, we developed a polarimetric imaging setup. In this case, the supporting plate was placed on a Vibraplane 5602-3672-31 imaging table (Kinetic Systems, Phoe nix, USA). The samples were illuminated with a XD-301 halogen light source (Alltion, Guangxi, China), placed 35 mm above the center of the sample plate. A FKB-VIS-40 band-pass filter (Thorlabs, New Jersey, USA) with central wavelength 400 nm was used to allow light with wavelengths comprised between 380 nm and 420 nm. A TRI050S-PC polarimetric camera (Lucid Vision Labs, Richmond, Canada) with a VS-LDA20 lens of focal length of 20 mm and maximum aperture of f/2.1 (Vital Vision, Paya Lebar, Singapore) was placed at a working distance of 90 mm from the sample supporting plate and with the viewing axis forming a certain angle θ (of 10° and 15°) with the plate surface, as illustrated in Fig. 2. The camera has a sensor with a total of 2448×2048 pixels arranged in units of four individual pixels, each one having a linearly polarized filter to register the intensities corresponding to polarization directions of 0°, 45°, 90° and 135° (as depicted in Fig. 2). Each of those units (of four polarized pixels) therefore defines an 'effective pixel' at the corresponding location in the image plane defined by the camera field of view. The values of irradiance registered at each direction of polarization at the location of each effective pixel are further combined -after demosaicing raw sensor file- the into four independent 1224×1024 pixel images (denoted as $I_0$, $I_{90}$, $I_{45}$ and $I_{135}$). Values of irradiance are further expressed normalized in 8 bits per pixel. The experiment was conducted in a dark room at ~24ºC and 50% relative humidity.

Polarization information was obtained from dry residues of each preparation. In doing so, 5-µL droplets were placed on a PTFE plate, also 1 mm in thickness, and approximately 50 mm × 20 mm of surface, with a spacing of about 8 mm to reduce potential crosstalk in light registered by each pixel of the imaging sensor. Each imaging preparation contained one droplet of each specimen for each type of dissolving fluid, that is, a total of four droplets in a row corresponding to G-LP, S-LP, control (viral culture medium), and pure fluid. An additional space was left in the row for delineating an empty region of the supporting plate equivalent (in image pixels) to a droplet, identified as background (BG). The fluid droplets were left to dry for at least 20 minutes.

The total number of samples analysed by imaging polarimetry was 475, corresponding to 380 fluid droplets and 95 equivalent areas of background. Table II details the number of samples (160 fluid droplets and BG areas) employed for 10-degree imaging and Table III those (315 fluid droplets and BG areas) for 15-degree imaging). The average number of pixels per fluid droplet and their corresponding standard deviation were 358±124 pixels for 10-degree imaging and 1281±581 pixels for 15-degree experiments. As described in a previous paper [34], a simplified disk model of a droplet imaged by a camera with an optical axis perpendicular to the plate yields an approximately equivalent pixel volume of 14 nL at 10 degrees and 4 nL at 15 degrees.

TABLE II
Number of samples used for imaging polarimetry at θ=10°

| Sample type | Concentrations | | | | | | | |
|---|---|---|---|---|---|---|---|---|
| | C1 | | C2 | | C3 | | C4 | |
| | P | N | P | N | P | N | P | N |
| G-LP in PBS | 4 | 4 | 4 | 4 | 4 | 4 | 4 | 4 |
| S-LP in PBS | 4 | | 4 | | 4 | | 4 | |
| G-LP in AS | 4 | 4 | 4 | 4 | 4 | 4 | 4 | 4 |
| S-LP in AS | 4 | | 4 | | 4 | | 4 | |
| PBS | 16 | | | | | | | |
| AS | 16 | | | | | | | |
| BG | 32 | | | | | | | |

Table II. Distribution of the 160 samples (128 fluid droplets and 32 droplet-equivalent background areas) used for imaging polarimetry at 10 degrees. Concentrations were C1 = 800 TU·µL$^{-1}$, C2 = 1500 TU·µL$^{-1}$, C3 = 3000 TU·µL$^{-1}$ and C4 = 4000 TU·µL$^{-1}$. G-LP = Lentiviral particles pseudotyped with the G protein of the vesicular stomatitis virus. S-LP = Lentiviral particles pseudotyped with the spike protein of the SARS-CoV-2. P = positive sample (solution with viral particles and culture medium), N= negative sample (solution with culture medium). PBS = phosphate buffered solution, AS = artificial saliva, BG = background (supporting plate).

## *2.7 Polarimetric imaging data analysis.*

The Stokes vector $S_r$ of the light reflected light by the sample in a given direction was obtained as described in equation (1) from the light intensity registered by the detector. As indicated, $I_0$, $I_{90}$, $I_{45}$ and $I_{135}$ denote the light intensity (irradiance) at the detector with the optical axis of the polarizers (analysers) set to 0, 90, 45 and 135 degrees, respectively. Fig. 2 shows the distribution of polarizer filters at each pixel of the camera sensor. Note also linearly polarized light was registered only, and therefore, $s_3$ was not determined in this analysis.

TABLE III
Number of samples used for imaging polarimetry at θ=15°

| Sample type | Concentrations | | | | | | | |
|---|---|---|---|---|---|---|---|---|
| | C1 | | C2 | | C3 | | C4 | |
| | P | N | P | N | P | N | P | N |
| G-LP in PBS | 8 | 8 | 8 | 8 | 8 | 8 | 8 | 8 |
| S-LP in PBS | 8 | | 8 | | 8 | | 8 | |
| G-LP in AS | 8 | 8 | 8 | 8 | 7 | 7 | 8 | 8 |
| S-LP in AS | 8 | | 8 | | 7 | | 8 | |
| PBS | 32 | | | | | | | |
| AS | 31 | | | | | | | |
| BG | 63 | | | | | | | |

Table III. Distribution of the 315 samples (252 fluid droplets and 63 droplet-equivalent background areas) used for imaging polarimetry at 15 degrees. Concentrations were C1 = 800 TU·µL$^{-1}$, C2 = 1500 TU·µL$^{-1}$, C3 = 3000 TU·µL$^{-1}$ and C4 = 4000 TU·µL$^{-1}$. G-LP = Lentiviral particles pseudotyped with the G protein of the vesicular stomatitis virus. S-LP = Lentiviral particles pseudotyped with the spike protein of the SARS-CoV-2. P = positive sample (solution with viral particles and culture medium), N= negative sample (solution with culture medium). PBS = phosphate buffered solution, AS = artificial saliva, BG = background (supporting plate).

$$S_r = \begin{bmatrix} s_0 \\ s_1 \\ s_2 \\ s_3 \end{bmatrix} = \begin{bmatrix} I_0 + I_{90} \\ I_0 - I_{90} \\ I_{45} + I_{135} \\ - \end{bmatrix} \quad (1)$$

The experiments of point polarimetry indicated that the DLP appears to be a good diagnostic tool for detecting contamination over the surface of interest. The DLP together with the angle of linear polarization (ALP) of the reflected beam are defined as functions of the Stokes parameters as:

$$DLP = \frac{\sqrt{s_1^2 + s_2^2}}{s_0} \qquad (2)$$

$$ALP = \frac{1}{2}\arctan\left(\frac{s_2}{s_1}\right) \qquad (3)$$

For the analysis of the polarimetric images, we defined normalized Stokes parameters as described in equation (4), where max($s_0$) denotes the maximum value of $s_0$ in the polarimetric image.

$$S_{r,n} = \begin{bmatrix} s_{0n} \\ s_{1n} \\ s_{2n} \\ s_{3n} \end{bmatrix} = \begin{bmatrix} s_0/\max(s_0) \\ s_1/s_0 \\ s_2/s_0 \\ - \end{bmatrix} \qquad (4)$$

Note that $s_{3n}$ is not determined in the employed set-up. In addition, we included three parameters, defined as relative differences between any combination of the Stokes parameters, as described in equations (5-7).

$$A_1 = \frac{s_1 - s_0}{s_1 + s_0} \qquad (5)$$

$$A_2 = \frac{s_2 - s_0}{s_2 + s_0} \qquad (6)$$

$$A_3 = \frac{s_1 - s_2}{s_1 + s_2} \qquad (7)$$

Using an in-house computer program for polarimetric image analysis developed in R2020b (The Mathworks Inc., Massachusetts, USA), droplet images were segmented. As indicated, each imaged plate contained four dry droplets (corresponding to G-LP in fluid, S-LP in fluid, control (viral culture medium) in fluid, and the pure fluid) and the additional background area. The corresponding values of the registered intensities for each direction of polarization ($I_0$, $I_{45}$, $I_{90}$, and $I_{135}$) and the 'polarimetric features' (namely, DLP, ALP, $S_{0n}$, $S_{1n}$, $S_{2n}$, $A_1$, $A_2$, and $A_3$) at each pixel were determined and stored.

### *2.8 Statistical analysis.*

Mean values of the DLP and ALP at each reflection angle (point measurements) were graphed for each type of viral particle and fluid, their corresponding negative controls and pure fluids, and the supporting plate. Droplet-averaged values of DLP, ALP, $S_{0n}$, $S_{1n}$, $S_{2n}$, $A_1$, $A_2$, and $A_3$ were calculated from the corresponding pixels for each sample from the polarimetric images, and their values and uncertainties were represented in boxplots. In addition, several combinations of polarimetric features were represented as two- and three-dimensional plots to explore their ability to discern among the different types of samples.

The t-test was performed on the arithmetic combinations of the parameters to distinguish the different samples under examination. Statistical significance was considered at the 95% confidence level. P-values below 0.0001 were considered as zero. Errors were reported here as the standard deviation.

### *2.9 Ethics and data availability.*

The study was approved by the Regional Research and Ethics Committee of University Hospitals 'Virgen Macarena and Virgen del Rocío', Seville, Spain (reference 0945-N-20, approved on 4/21/2020). Data supporting this work are available from the authors under reasonable request.

## 3. Results

It is known that biomolecules can alter the state of polarization of the scattered light [45]. In order to explore whether this phenomenon can be exploited for the intended detection and quantification of viruses, we first determined the optimum reflection angle that maximized the DLP of the light reflected by the (dry) fluid samples. Then, the polarimetric images were obtained at the optimum viewing angle. A set of numerical features describing the polarization of the light reflected off the samples was numerically calculated for the image pixels corresponding to each sample and analyzed.

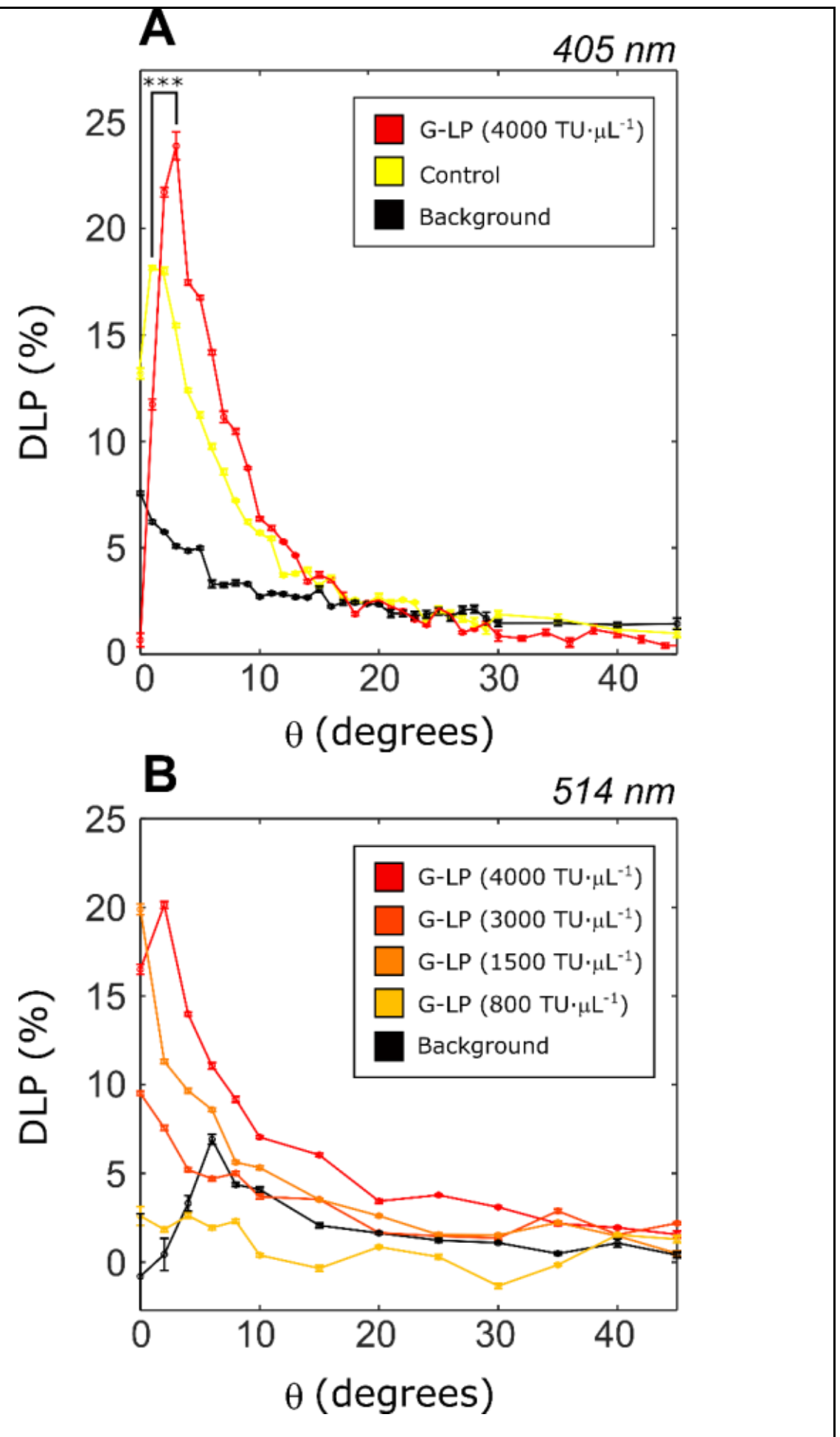

Fig. 3. Measured DLP curves as functions of the viewing angle (θ) at 405 nm (Panel A) and 514 nm (Panel B) of dry residues of preparations of lentiviral particles pseudotyped with the G protein of the vesicular stomatitis virus (G-LP) and the corresponding negative controls in four different concentrations in artificial saliva. Each data point (solid symbols) correspond to average value of 1000 measurements obtained in 1 second. The error bars denote the standard deviation. The background represents the DLP obtained from the supporting (clean) PTFE plate. ***p-value < 0.0001 from two-tailed Wilcoxon rank-sum test.

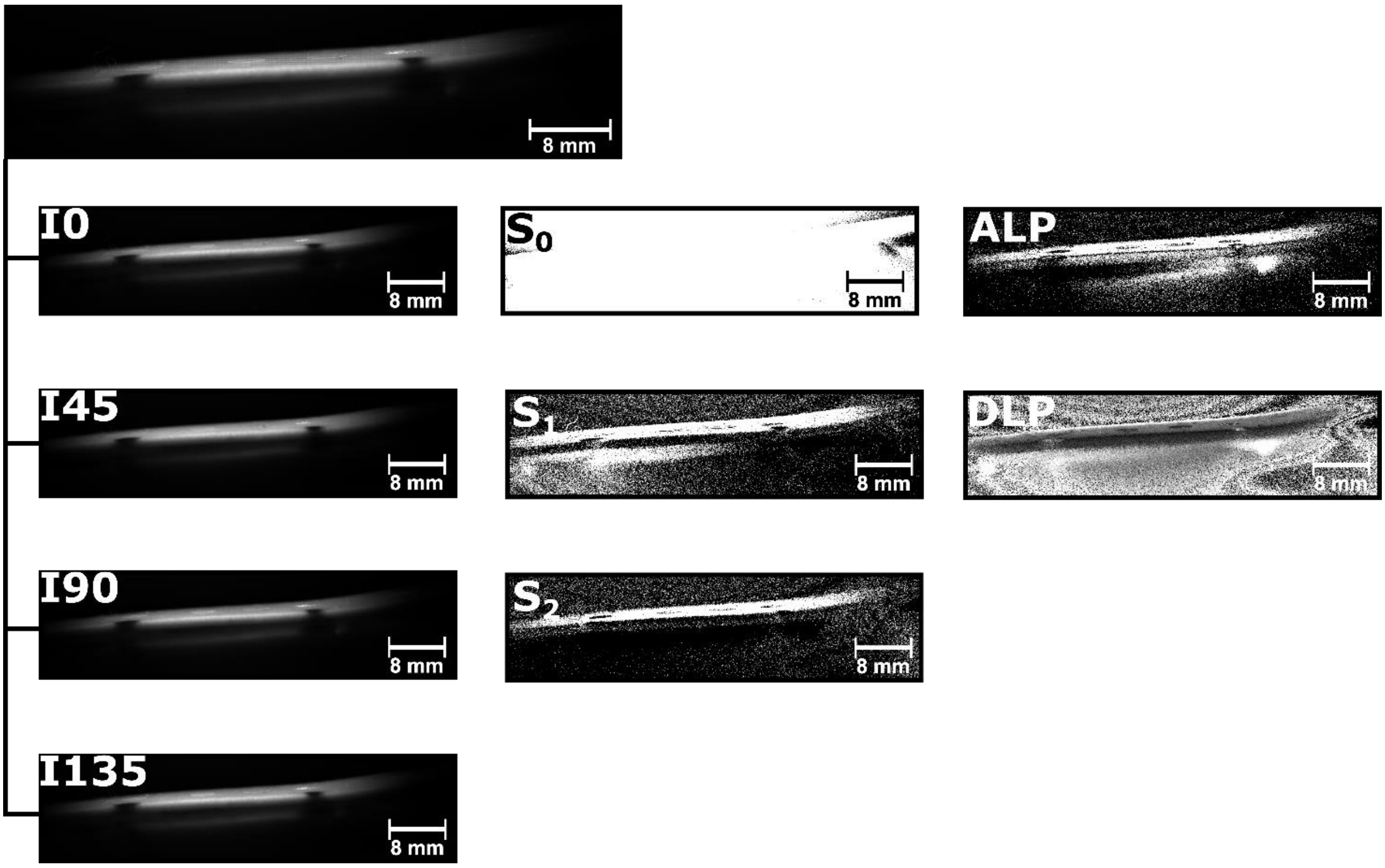

Fig. 4. Polarimetric images obtained with the camera placed in a 10-degree configuration in artificial saliva. Top image is the luminance image of the sample. The four (dry) fluid droplets correspond (from left to right) to G-LP, S-LP, control, and pure fluid. Images I0, I45, I90 and I135 were obtained after demosaicing and correspond to irradiance of the 0-degree, 45-degree, 90-degree and 135-degree polarization components. Images $s_0$, $s_1$ and $s_2$ were obtained by computing the Stokes parameters at each pixel. Likewise, images corresponding to the angle and degree of linear polarization (ALP and DLP) were computed.

### *3.1 Determination of the angle of maximal DLP.*

The objective of the point polarimetry experiments was finding the optimum illumination and observing geometry, and the corresponding viewing angle θ (i.e., the complementary of the reflection angle) at which the DLP of the reflected light was maximum, as this allows for maximizing the information obtained from the polarization properties of the sample under examination. Fig. 3A shows the value of the DLP obtained from a preparation of G-LP in artificial saliva at the highest concentration and its negative control (i.e., samples with fluid and culture medium but without viral particles).

There were statistically significant differences between the maximum value of the DLP in both cases ($p$-*value*=0.00). The angle thus obtained was θ = 3° at 405 nm. This angle remained constant for all other viral concentrations. Likewise, the angle that maximized the DLP for the control sample was 1°. In both cases, the DLP decreased as the viewing angle increased. The measured values in both cases tend to the values obtained for clean PTFE plate (background) for θ > 20°. As shown in Fig. 3, panel B, similar results are obtained at 514 nm. The value of the DLP showed statistically significant differences ($p$-*value*=0.00) among the viral concentrations. These differences were larger at lower viewing angles. Note also that, at this wavelength, the sample plate exhibits a peak at 6 degrees, as shown in Fig. 3B.

### *3.2 Viral detection using polarimetric imaging.*

To obtain redundant information simultaneously from many points of a given sample preparation we used a polarimetric camera placed to image the samples at a certain angle. However, the viewing direction of the camera must balance two opposing factors: the small value of maximal DLP determined by point polarimetry (with respect to the supporting plate), which would define an approximately tangential imaging acquisition, and the need to obtain 'large enough' images of the sample droplets to provide a sufficient number of pixels per droplet for numerical assessment. When the camera is close to the plane defined by the sample supporting plate, the distortion and parallax of the imaged field of view increase notably. Besides the geometric distortion of perspective, it must be taken into account that the sample volumes corresponding to each image pixel also differ.

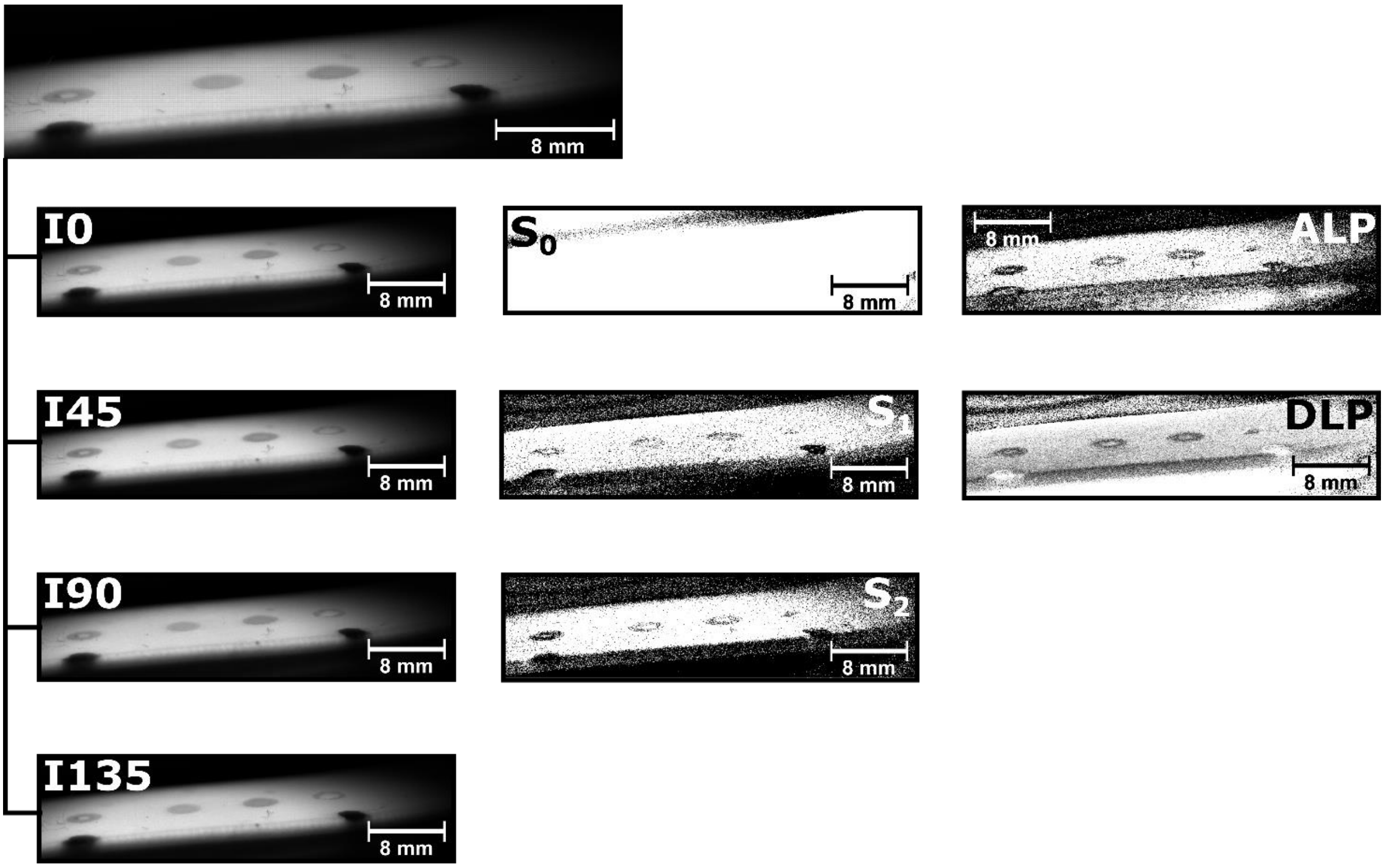

Fig. 5. Polarimetric images obtained with the camera placed in a 15-degree configuration. Top image is the luminance image of the samples in artificial saliva. The four (dry) fluid droplets correspond (from left to right) to G-LP, S-LP, control, and pure fluid. Images I0, I45, I90 and I135 were obtained after demosaicing and correspond to irradiance of the 0-degree, 45-degree, 90-degree and 135-degree polarization components. Images $s_0$, $s_1$ and $s_2$ were obtained by computing the Stokes parameters at each pixel. Likewise, images corresponding to the angle and degree of linear polarization (ALP and DLP) were computed.

Images were therefore obtained at $\theta = 10°$ and $15°$, as shown in Fig. 4 and Fig. 5, respectively. Note these angles do not maximize the value of DLP as described previously for 405 nm. However, considering that the light source was band-pass filtered between 380 nm and 420 nm, and that the (centre) axis of illumination was perpendicular to the sample supporting plate, these observation angles provided sufficient information about the vectorial properties of the scattered light. As mentioned in the methodology, the Stokes parameters $s_0$, $s_1$ and $s_2$ were obtained from the light intensities captures by the camera in the different linear polarization angles. Next, the DLP and the ALP were computed. Images built by calculating all these two parameters pixel by pixel are displayed in Fig. 4 and Fig. 5. Which also show similar images constructed from the other parameters used in this analysis, that is, the normalized Stokes parameters $s_{0n}$, $s_{1n}$ and $s_{2n}$, and the relative differences defined ($A_1$, $A_2$ and $A_3$).

## 4. Discussion

The spread of the SARS-CoV-2 virus is an obvious cause of severe concern worldwide. Despite the high vaccination rates achieved in many territories, the transmission of new variants of the virus is raising fears, as these strains incorporate mutations in the spike protein able to reduce the affinity of existing antibodies [56, 57]. Therefore, continued, massive testing of the population remains as one of the principal measures of public health to contain the spread of the pathogen. Along these lines, numerous molecular technologies have been developed for the detection of the genetic material of the virus [58], viral antigens [59], or the specific immune response triggered after the infection [60]. In addition, optical technologies also boost SARS-CoV-2 detection capabilities by providing new methods able to identify the presence of the virus [35, 61]. Recent works include the analysis of hyperspectral images of diffuse reflectance for the detection and quantification of synthetic viral models [32] and of SARS-CoV-2 in human samples [34] reported by the authors.

It is important to note that image-based virus detection schemes allow for rapid analysis of multiple samples (all droplets included within the field of view of the imaging system), using a reagent-free, relatively simple and safe (non-contact) configuration suitable for use by minimally trained operators even in resource-constrained settings. The results reported here on the differentiation between fluid samples containing two similar viral models of SARS-CoV-2 using polarimetric imaging analysis can be incorporated along with other existing technologies for making optical detection more robust.

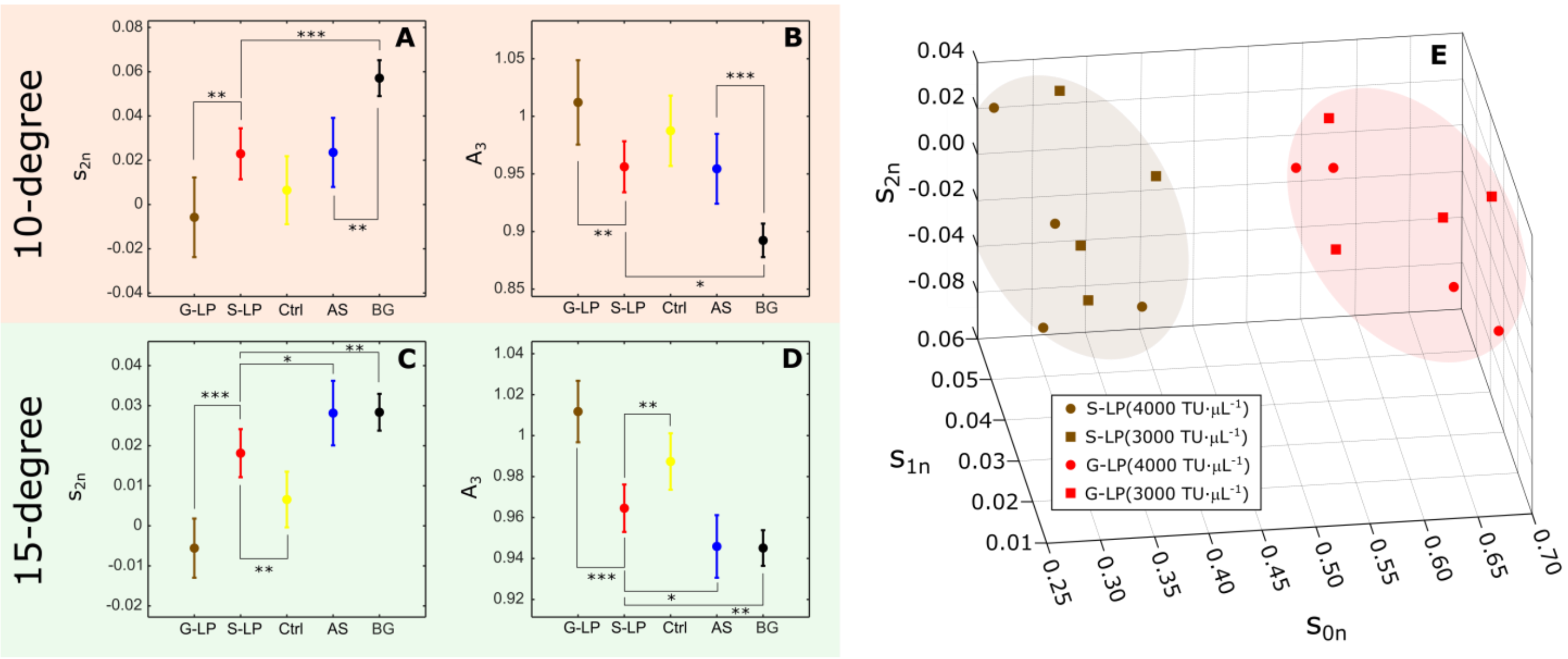

Fig. 6. Polarimetric parameters. Panels **A** and **B** show the mean values of the $s_{2n}$ and $A_3$ parameters respectively obtained with the polarimetric camera in a 10° configuration. Panels **C** and **D** show the mean values of the $s_{2n}$ and $A_3$ parameters respectively obtained with the polarimetric camera in a 15° configuration. Error bars denote the calculated uncertainties. *p-value<0.1, **p-value<0.05, ***p-value < 0.005 from t-test (n=4). These parameters were obtained from samples containing lentiviral particles pseudotyped with the G protein of the vesicular stomatitis virus (G-LP) and lentiviral particles pseudotyped with the spike protein of the SARS-CoV-2 (S-LP) prepared at 4000 TU·µL$^{-1}$ concentration. Negative controls (Ctrl) were obtained from non-transfected culture media in the same concentrations. All biological specimens were re-suspended in artificial saliva (AS). The background (BG) given by the supporting plate was also characterized. Panel **E** shows a 3-dimensional scatter plot of the mean values of the normalized Stokes parameters $s_{0n}$, $s_{1n}$ and $s_{2n}$ obtained from viral preparations at high concentrations (4000 TU·µL$^{-1}$ and 3000 TU·µL$^{-1}$) in phosphate buffered solution obtained in a 10-degree imaging configuration.

We first explored whether there were differences between the DLP of the light reflected by samples containing one type of viral particles comparable to the SARS-CoV-2 virus (i.e., G-LP) and their negative controls. Certainly, the differences found revealed that these differences were more evident at higher concentrations. Note that the liquid samples were left to dry, and therefore, these differences were caused by the composition of the residues, in this case, the presence or not of lentiviral particles. Therefore, the molecular composition and concentration of the G-LPs was successfully detected using repeated DLP point measurements of the reflected light at a certain angle range at two wavelengths. This result agrees with the aforementioned detection and quantification of the same type of samples, in the same biofluids and concentrations (both as liquid droplets and dry residues) using hyperspectral image analysis [34].

The next step was to compare among samples with the two similar viral particles, G-LP and S-LP, being the minimal difference among them the presence of the characteristic spike protein of the SARS-CoV-2. For that purpose, we developed the described polarimetric imaging set-up with θ within the range previously determined from the point measurements. This imaging approach also possesses several limitations: (1) uneven illumination over the whole sample as a consequence of using conical illumination, (2) a relatively wide bandwidth of the light source, (3) crosstalk from reflected light at different sample points, (4) differences in the illumination angles caused by the rugosity of the supporting plate, (5) image distortion and aberrations produced by the lens of the camera, or (6) parallax distortion due to the small imaging angle among others. However, the advantages of the proposed methodology overcome its limitations, as imaging allows for the simultaneous acquisition of information from many samples (i.e., suitable for massive test implementations) and from many points (pixels) of each sample (i.e., providing data suitable for enhanced extraction of information). In addition, we propose the combined use of multiple parameters for a precise numerical analysis of the differences in the state of polarization of the scattered light which are expected to allow for effective discernment of different virions in the samples (and their absence in the corresponding negative controls). Unlike other applications of polarimetric imaging, each sample here was characterized by an extended set of parameters that quantify, directly or indirectly, vectorial properties of the diffusedly reflected light. Those parameters include the commonly used irradiances at each direction of the analyzer optical axis ($I_0$, $I_{90}$, $I_{45}$ and $I_{135}$), the (original and normalized) Stokes parameters ($s_o$, $s_1$, $s_2$, $s_{on}$, $s_{1n}$, $s_{2n}$) and their derived magnitudes (DLP, ALP) plus the newly introduced relative differences ($A_1$, $A_2$ and $A_3$). Furthermore, as indicated, the proposed imaging approach simplifies the experimental setup and reduces the level of training required to perform a test. Unfortunately, to exploit the information contained in the polarization state of the scattered light, a low imaging angle is required. While differences in the polarization properties of the light could be maximized at the 5° imaging configuration, the reduced dimensions of the resulting sample images prevent from obtaining redundant information from larger groups of pixels. Note that even though samples with G-LPs and S-LPs could be distinguished at a 10-degree configuration, differentiation from the negative control was not possible using one parameter alone. Nevertheless, differentiation among the two viral species and the negative control was successful using a 15° imaging configuration. This can be explained in terms of substantially lower sample areas covered by each pixel in the image. In addition, in said imaging configuration, the fluid medium (AS) and the supporting background material (PTFE) produced similar changes in the polarization parameters under study that hampered their differentiation. Therefore, the alterations of the polarization state of the incident light produced by the biological content were sufficient for differentiating dried fluid samples containing two similar viral particles, one expressing the G protein of the VSV, and the other, the S protein of the SARS-CoV-2, both within the same lentiviral system.

As presented here, these differences can be enhanced if different parameters are combined and processed using multivariate analysis. Although there is an evident need for extending these results beyond the use of laboratory engineered viruses, we anticipate that the biological signature of viral pathogens, that is, the molecular composition that allows for their detection, embeds certain 'information' in the reflected light. Note, for example, that previous studies on imaging spectroscopy [32, 34] found important characteristic features at similar wavelengths. Notably, the results obtained here agree with the aforementioned numerical simulations [48] in which changes in the circular polarization of ultraviolet light (310 nm) allowed for the discernment of computational viral models similar to the ones employed in our experiments.

In this work we explored the use of this technology for identifying potentially contaminated surfaces containing dry residues. A supporting plate containing dry samples was used as a model of a fomite. As in the case of hyperspectral imaging -and, possibly, combined with it-, this approach could also be exploited for mass screening of respiratory diseases, including COVID-19. In doing so, studies that include human specimens should be conducted to determine the sensitivity and specificity of the technology. If properly linked to other approaches, optical detection of pathogens by a reagent-free, non-contact, easy-to-implement approach has the potential to provide truly useful tools to face biosecurity hazards, as it would allow for scanning of multiple samples within very short time frames, opening its use to transport hubs or mass events.

## 5. Conclusions

In summary, we present a proof-of-concept study on the use of polarimetric image analysis for detection and discernment of synthetic models of SARS-CoV-2 in dry fluid droplets under visible light. Experiments were conducted using two types of engineered viral models (lentiviral particles with and without Spike protein), in two fluids (saline solution and artificial saliva), in four concentrations. Mueller polarimetry of 120 samples determined the optimal angles for polarimetric imaging. Changes in the linear polarization of light diffusely scattered by other 475 samples were quantified in a per-pixel approach. Obtained results agree with computational simulations in related digital models by other authors. Polarimetric imaging shows a high potential for non-contact, reagent-free simultaneous analysis of multiple fluid samples. Further studies with human samples are required, as this approach could be combined with other optical techniques for cost-effective, relatively easy-to-implement virus screening including SARS-CoV-2.

## Acknowledgements

This study was funded by grants number COV20-00080 and COV20-00173 of the 2020 Emergency Call for Research Projects about the SARS-CoV-2 virus and the COVID-19 disease of the Institute of Health 'Carlos III', Spanish Ministry of Science and Innovation, and by grant number EQC2019-006240-P funded by MICIN/AEI/10.13039/501100011033 and by "ERDF A way of making Europe". ABR was supported by grant number RTI2018-094465-J-I00 funded by MICIN/AEI/10.13039/501100011033 and by "ERDF A way of making Europe". The work of OM, JCGM, and JLR has been partially funded by grant LEONIDAS (RTI2018-095330-B-100), and the Center of Excellence Severo Ochoa award to the Instituto de Astrofísica de Andalucía (SEV-2017-0709). This work has been supported by the European Commission through the Joint Research Center (JRC) HUMAINT project.

The authors would like to gratefully acknowledge the assistance of the members of the Explosive Ordnance Disposal – Chemical, Biological, Radiological & Nuclear (EOD-CBRN) Group of the Spanish National Police, whose identities cannot be disclosed, and who are represented here by JMNG.